\documentclass[aps,superscriptaddress,amsmath,amssymb,floatfix,twocolumn,showpacs,amsfonts,longbibliography]{revtex4-1}
\usepackage{times}
\usepackage[varg]{txfonts}
\usepackage{textcomp}
\usepackage{graphicx}
\usepackage{subfigure}
\usepackage{tabu}
\usepackage{color}
\usepackage[colorlinks=true,citecolor=blue,urlcolor=blue,linkcolor=blue,hyperindex]{hyperref}
\usepackage{braket}
\usepackage{float}
\usepackage{overpic}
\usepackage{amssymb}

\allowdisplaybreaks

\usepackage{times}

\def\be{\begin{equation}} \def\ee{\end{equation}}
\def\bea{\begin{eqnarray}} \def\eea{\end{eqnarray}}

\def\bk{{\bf k}}

\def\be{{\bf e}}

\def\rw{\rightarrow}

\begin{document}

\title{Majorana corner and hinge modes in second-order topological \\ insulator-superconductor heterostructures}
\author{Zhongbo Yan}
\email{yanzhb5@mail.sysu.edu.cn}
\affiliation{ School of Physics, Sun Yat-Sen University, Guangzhou 510275, China}


\begin{abstract}
As platforms of Majorana modes,  topological insulator (quantum anomalous Hall insulator)/superconductor (SC) heterostructures have
attracted tremendous attention over the past decade. Here we substitute the topological insulator by its higher-order counterparts.
Concretely, we consider second-order topological insulators (SOTIs) without time-reversal symmetry and investigate
SOTI/SC heterostructures in both two and three dimensions. Remarkably, we find that such novel heterostructures provide
natural realizations of second-order topological superconductors (SOTSCs) which host Majorana corner modes in two dimensions
and chiral Majorana hinge modes in three dimensions. As here the realization of SOTSCs  requires neither special pairings
nor magnetic fields,  such SOTI/SC heterostructures are outstanding platforms of Majorana modes and may have wide applications in future.

\end{abstract}

\maketitle

Over the past decade, topological superconductors (TSCs) have attracted continuous and
tremendous attention\cite{qi2011,alicea2012new,Beenakker2013,stanescu2013majorana, leijnse2012introduction,Elliott2015,sarma2015majorana,sato2016majorana,aguado2017}.
Among various TSCs, one-dimensional ($1d$) and two-dimensional ($2d$) TSCs
without time-reversal symmetry (TRS) have attracted particular interest as they harbor
Majorana zero modes (MZMs) at their boundaries\cite{kitaev2001unpaired,oreg2010helical,lutchyn2010} and in the cores of vortices\cite{read2000,fu2007c,sau2010,alicea2010}, respectively.
Owing to their fractional nature, MZMs are ideal candidates
to construct nonlocal qubits immune to local decoherence\cite{kitaev2001unpaired}.
Moreover, owing to their non-Abelian statistics\cite{ivanov2001},
their braiding operations are found to realize elementary quantum gates.
Thus, MZMs are believed to be building blocks of topological quantum computation\cite{nayak2008} and
have been actively sought in experiments\cite{mourik2012signatures, rokhinson2012fractional,
deng2012anomalous, das2012zero,finck2013,nadj2014observation,albrecht2016exponential,Deng2016Majorana,
zhang2018quantized,Sun2016Majorana,wang2018evidence}.

As is known,  odd-parity superconductors (SCs) provide natural realizations of TSCs, however,
they are unfortunately rare in nature. In a seminal paper\cite{fu2007c}, Fu and Kane
pointed out that  topological insulator (TI)/SC heterostructures provide
an effective realization of odd-parity superconductivity.  Accordingly, in the presence of
magnetic field, vortices emerging in such heterostructures are found to carry MZMs.
In a later influential paper\cite{qi2010chiral}, Qi \textit{et al} pointed out that  quantum anomalous Hall insulator (QAHI)/SC
heterostructures provide a simple realization of $2d$ chiral TSCs which
harbor not only vortex-core MZMs, but also chiral Majorana edge modes.
These two theoretical works have triggered a lot of experimental works
on TI(QAHI)/SC heterostructures\cite{Sun2016Majorana,wang2018evidence,wang2012coexistence, Wang2013proximity,Zareapour2012proximity,Xu2015MZM,lv2016,zhang2018iron,Liu2018MZM,He2017chiral,Kayyalha2019chiral,chen2019quantized,Zhu2019MZM},
and remarkable progress in detecting
vortex-core MZMs has been witnessed in recent years\cite{Sun2016Majorana,wang2018evidence,chen2019quantized,Zhu2019MZM}.

Very recently, TIs and TSCs have been generalized to include their higher-order counterparts\cite{benalcazar2017quantized,Song2017higher,Langbehn2017hosc,Benalcazar2017prb,Schindler2018HOTIa,ezawa2018higher,Rasmussen2018HOSPT,You2018hospt,
Khalaf2018hosc,Geier2018hosc,Franca2018HOTI,Roy2019HOTI,Trifunovic2019HOTI,Ahn2018HOTI,Kudo2019HOTMI}.
Importantly, higher-order TIs (HOTIs) and TSCs (HOTSCs) have extended the conventional
bulk-boundary correspondence. Accordingly,  an $n$-th order TI or TSC in $d$ dimensions host
($d-n$)-dimensional boundary modes. For instance, a second-order TI (SOTI) in $2d$ and $3d$
host zero-dimensional ($0d$) corner modes and $1d$ hinge modes, respectively. The existence of
HOTIs and the lessons from the study of TI(QAHI)/SC heterostructures lead us to ask the natural
question that whether  Majorana corner modes (MCMs, i.e., MZMs bound at the corners) or chiral Majorana hinge modes (CMHMs)
can also be achieved in a HOTI/SC heterostructure.
It is worth noting that such a question is quite timely as recently the electronic material candidates
for SOTIs, both in two dimensions ($2D$) and three dimensions ($3D$), are growing\cite{schindler2018HOTI,yue2019symmetry,Wang2018XTe,Xu2019HOTI,Sheng2019SOTI,Lee2019HOTI,Chen2019HOTI}.
Moreover, signature of MZM has also been observed in a heterostructure which consists of a
bismuth thin film (a SOTI with TRS\cite{schindler2018HOTI}), a conventional $s$-wave SC,
and magnetic iron clusters\cite{Jack2019observation}.

In this work, we consider SOTIs without TRS and  investigate SOTI/SC heterostructures in both $2D$ and $3D$.
Remarkably, we find that such heterostructures provide natural realizations of second-order
topological superconductors (SOTSCs) which host MCMs
in $2D$ and CMHMs in $3D$. Furthermore, here the realization of SOTSCs  does not require the pairing
of SCs  to take any specific form. It can be achieved for both unconventional
SCs and conventional $s$-wave SCs. In addition, it does not need magnetic fields
or the deposition of magnetic atoms. In comparison to previous
proposals\cite{MZM-Hopf,Shapourian2018sotsc,Zhu2018hosc,Yan2018hosc,Wang2018hosc,Wang2018hosc2,Liu2018hosc,Hsu2018,Pan2018SOTSC,
Bultinck2019HOSC,Yang2019hinge,Volpez2019SOTSC,Wu2019hosc,Zeng2019hosc,Ghorashi2019HOSC,Zhang2019hinge,Kheirkhah2019SOTSC,
Hsu2019HOSC,Wu2019swave,Laubscher2019FHOSC,Zhang2019HOTSC,Zhu2019mixed,Zhang2019hoscb,Wu2019hoscb,Yan2019hoscb,Ahn2019hosc}, these merits make SOTI/SC heterostructures stand out, and potentially allow them to have wide applications in topological quantum computation\cite{You2018hosc,Bomantara2019braiding,Pahomi2019braiding}.

\begin{figure}
\subfigure{\includegraphics[width=7cm, height=2.6cm]{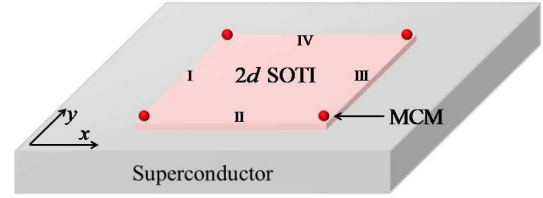}}
\caption{Schematic illustration. A $2d$ SOTI is grown on
a superconductor which can be either conventional or unconventional. Each corner of the SOTI
host one MZM. }  \label{sketch}
\end{figure}

{\it MCMs in a  $2d$ SOTI-SC heterostructure.---} A $2d$ SOTI/SC heterostructure (Fig.\ref{sketch}) could be described by a  Bogoliubov-de
Gennes (BdG) Hamiltonian, $H=\sum_{\bk}\Psi_{\bk}^{\dag}H(\bk)\Psi_{\bk}$,
with $\Psi_{\bk}=(c_{a,\bk\uparrow},c_{b,\bk\uparrow}, c_{a,\bk\downarrow},c_{b,\bk\downarrow},
c_{a,-\bk\uparrow}^{\dag},c_{b,-\bk\uparrow}^{\dag}, c_{a,-\bk\downarrow}^{\dag},c_{b,-\bk\downarrow}^{\dag})^{T}$ and
\begin{eqnarray}
H(\bk)&=&\epsilon(\bk)\sigma_{z}\tau_{z}+\lambda_{x}\sin k_{x}\sigma_{x}s_{z}+\lambda_{y}\sin k_{y}\sigma_{y}\tau_{z}\nonumber\\
&&+\Lambda(\bk)\sigma_{x}s_{x}\tau_{z}+\mu\tau_{z}+\Delta(\bk)s_{y}\tau_{y},\label{minimal}
\end{eqnarray}
where $\sigma_{i}$, $s_{i}$ and $\tau_{i}$ are Pauli matrices
in orbit $(a,b)$, spin ($\uparrow,\downarrow$) and particle-hole spaces, respectively;
$\epsilon(\bk)=m_{0}-t_{x}\cos k_{x}-t_{y}\cos k_{y}$ is the kinetic energy;
$\Lambda(\bk)=\Lambda_{x}\cos k_{x}-\Lambda_{y}\cos k_{y}$ is a TRS breaking term
crucial for the realization of SOTI;
$\mu$ is the chemical potential, and $\Delta(\bk)=\Delta_{0}+\Delta_{x}\cos k_{x}+\Delta_{y}\cos k_{y}$ represents
the pairing. Such a form  is general enough to model $s$-wave, $s_{\pm}$-wave and $d$-wave pairings\cite{Yan2018hosc}.
For  convenience, the lattice constants have been set to unit, and $t_{x,y}$, $\lambda_{x,y}$ and $\Lambda_{x,y}$ are set to be positive throughout
this work.

Let us focus on the normal state first. Without the terms in the second line of Eq.(\ref{minimal}),
the Hamiltonian describes a $2d$ first-order TI when $\prod_{\alpha,\beta=\pm1}(m_{0}+\alpha t_{x}+\beta t_{y})<0$\cite{fu2007a}.
Accordingly, when open boundary condition is taken, gapless helical modes will appear on the boundary.
Adding the $\Lambda(\bk)$ term breaks TRS and consequently gaps out the helical modes, resulting in a transition
from a first-order TI to a SOTI. When open boundary conditions are taken in both the $x$ and $y$ directions,
one can find that in the SOTI phase, each corner of the system will harbor one zero-energy bound state with a fractional charge
$e/2$\cite{Song2017higher}. The pinning of  the corner modes' energy to zero is due to the existence of a chiral symmetry
(the operator is $\sigma_{x}s_{y}$. When superconductivity enters, the operator
is accordingly modified as $\sigma_{x}s_{y}\tau_{z}$). However, this chiral symmetry is just an accidental symmetry,
adding an arbitrary term proportional to the identity matrix (e.g., the chemical potential) immediately
breaks this symmetry and accordingly shifts the energy away from zero.
Nevertheless, whether the chiral symmetry is preserved or not does not affect our following discussions
since the particle-hole symmetry of a SC is sufficient to guarantee the topological robustness
of MCMs.

\begin{figure*}[t!]
\subfigure{\includegraphics[width=5cm, height=4.5cm]{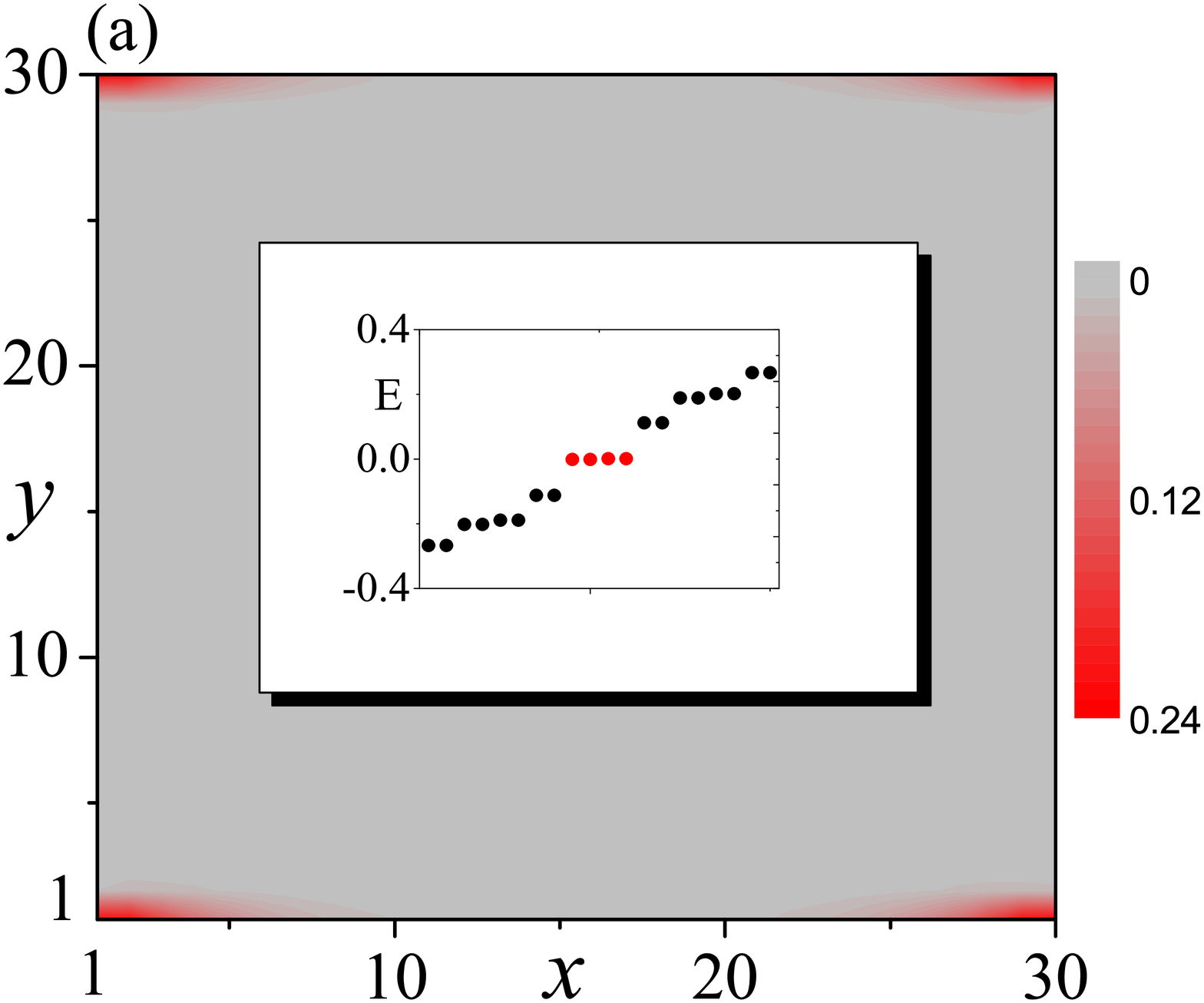}}
\subfigure{\includegraphics[width=5cm, height=4.5cm]{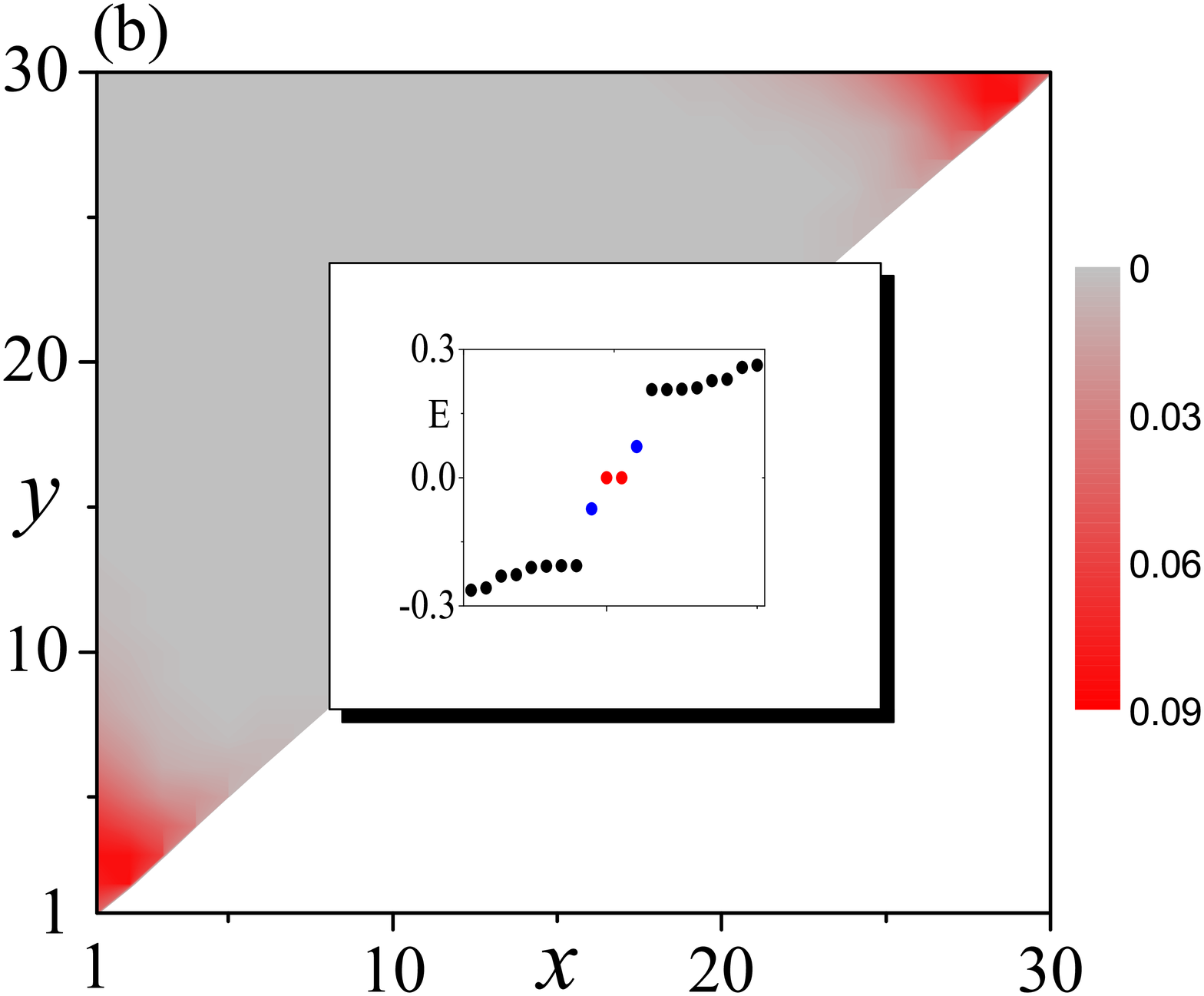}}
\subfigure{\includegraphics[width=5cm, height=4.5cm]{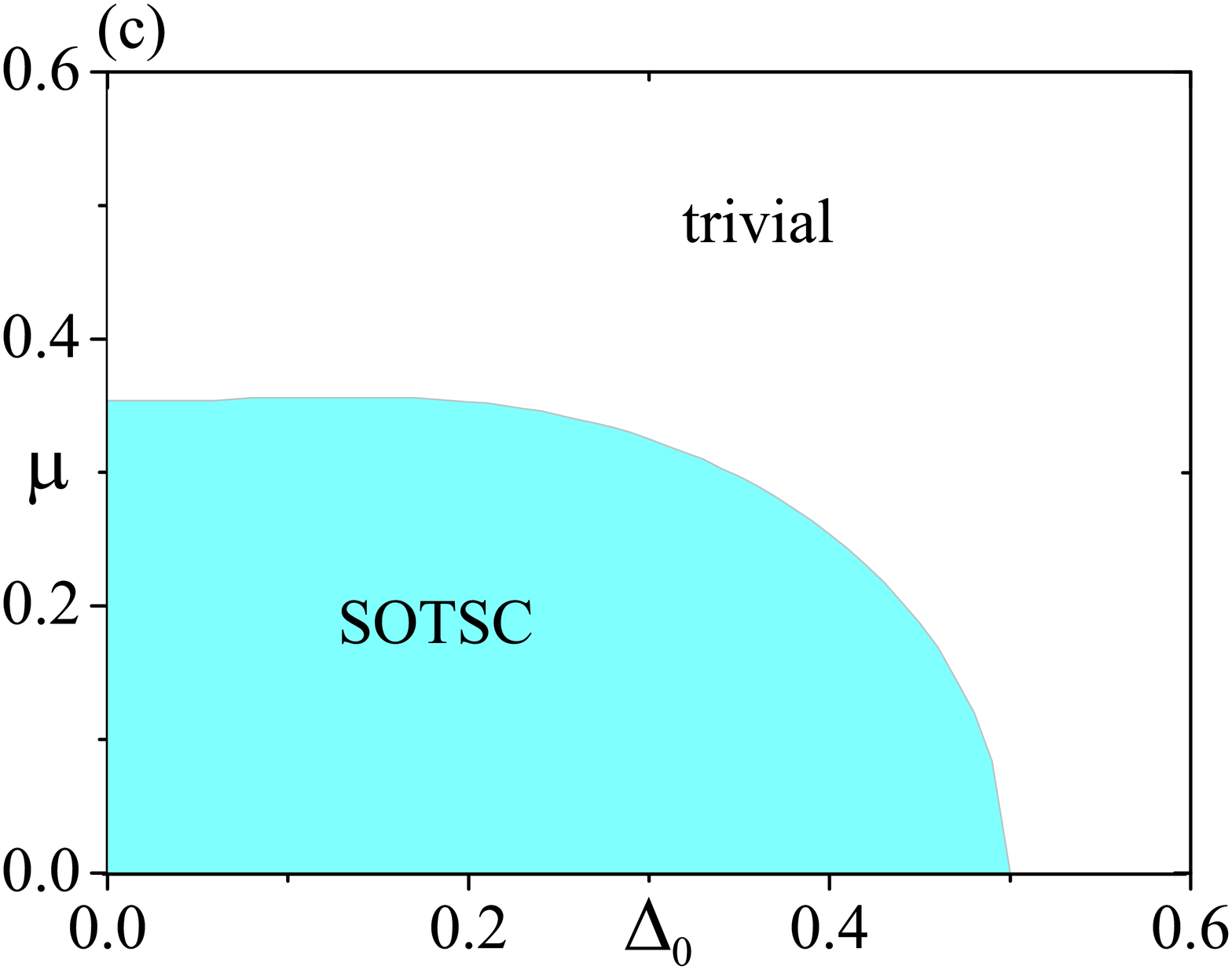}}
\caption{ MCMs in a SOTI/s-wave SC heterostucture. Common parameters:
$t_{x}=1$, $\lambda_{x}=\lambda_{y}=\Lambda_{x}=\Lambda_{y}=1.0$,  and $\Delta_{x}=\Delta_{y}=0$. (a) The wave function profiles of MCMs in a square sample.
We take $t_{y}=0.5$ to make the hoppings anisotropy. Furthermore,  we take $\mu=0.1$ to break the chiral symmetry, meanwhile, we take $m_{0}=1.0$ and  $\Delta_{0}=0.6$, so that
$m=t_{x}+t_{y}-m_{0}=0.5$ and the criterion $m\Lambda_{x}/t_{x}<\Delta_{0}<m\Lambda_{y}/t_{y}$ is satisfied.  The inset shows energies near zero.
There are four MZMs, with one per corner.
(b) The wave function profiles of MCMs in an isosceles-right-triangle sample.
We take $t_{y}=1.0$, $m_{0}=1.5$, and $\Delta_{0}=0.2$. Accordingly,  the criterion
$m\Lambda_{x}/t_{x}<\Delta_{0}<m\Lambda_{y}/t_{y}$ is not fulfilled, but $0<\Delta_{0}<m\Lambda_{x}/t_{x}$ is satisfied.
The numerical results reveal
that there are two MZMs whose wave functions are localized
around the two $\pi/4$-angle corners, agreeing with the edge theory.
The two modes labeled by blue dots in the inset
correspond to bound states located at the $\pi/2$-angle corner.
Their energies are not pinned to zero  since a finite $\mu$ breaks the chiral symmetry.
(c) A representative phase diagram for the isosceles-right-triangle geometry. $t_{y}=1$. ``trivial'' denotes region within which
MCMs are absent.}  \label{twomcm}
\end{figure*}

To see the effect of superconductivity intuitively, let us focus on the case with chiral symmetry first.
As is known, when a chiral electronic mode is in proximity to a SC,  it  becomes two chiral Majorana
modes in the weak-pairing limit\cite{qi2010chiral}. Similarly, when a SOTI is in proximity to a SC,
each charged corner mode will become two MCMs. However, as the wave functions of the two MCMs overlap in space,
they are not robust against local perturbations and disorders.
Therefore, one may naively think that robust MZMs in general can not be realized in a SOTI/SC heterostructure.
However, the simple ``one-to-two'' picture above is only valid in the weak-pairing limit.
In Fig.\ref{twomcm}(a) , we have shown explicitly that for a sample with square geometry, when the pairing amplitude exceeds some critical value,
a SOTSC phase with four well-separated MCMs can be achieved  even though the pairing of SC is $s$-wave.
As now each corner  has only one MZM, the particle-hole symmetry guarantees that
these MCMs are robust against local perturbations, as well as random disorders if
the disorder strength is weaker than some critical value.

{\it Edge theory.---} To see how the SOTSC phase and the concomitant MCMs are  realized,
we investigate the edge theory for an intuitive understanding. For simplicity, we still focus on the $\mu=0$ case
and consider the continuum model corresponding to a low-energy expansion of the lattice
Hamiltonian in Eq. (\ref{minimal}) to second order around $\bk=(0,0)$:
\begin{eqnarray}
H(\bk)&=&(\frac{t_{x}}{2}k_{x}^{2}+\frac{t_{y}}{2}k_{y}^{2}-m)\sigma_{z}\tau_{z}+\lambda_{x}k_{x}\sigma_{x}s_{z}+\lambda_{y}k_{y}\sigma_{y}\tau_{z}\nonumber\\
&&+\Lambda(k_{x},k_{y})\sigma_{x}s_{x}\tau_{z}
+\Delta(k_{x},k_{y})s_{y}\tau_{y},
\end{eqnarray}
where $m=t_{x}+t_{y}-m_{0}$,  $\Lambda(k_{x},k_{y})=\Lambda-\frac{1}{2}(\Lambda_{x} k_{x}^{2}-\Lambda_y k_{y}^{2})$
with $\Lambda=\Lambda_{x}-\Lambda_{y}$, and $\Delta(k_{x},k_{y})=\Delta-\frac{1}{2}(\Delta_{x} k_{x}^{2}+\Delta_y k_{y}^{2})$ with $\Delta=\Delta_{0}+\Delta_{x}+\Delta_{y}$. We consider $m>0$ so that without the terms in the second line the Hamiltonian
describes a first-order TI. For convenience, we label the four edges of a square sample I, II, III,
and IV (see Fig. \ref{sketch}) and define a $1d$ “boundary coordinate” $l$ which grows in a counterclockwise fashion along the edges.
Let us focus on the edge (I) first. To obtain the corresponding low-energy Hamiltonian which describes the edge modes,  we perform the replacement
$k_{x}\rw -i\partial_{x}$ and decompose the Hamiltonian into two parts\cite{Yan2018hosc}, i.e., $H=H_{0}+H_{p}$ with
\begin{eqnarray}
H_{0}(-i\partial_{x})&=&(-\frac{t_{x}}{2}\partial_{x}^{2}-m)\sigma_{z}\tau_{z}-i\lambda_{x}\partial_{x}\sigma_{x}s_{z},\nonumber\\
H_{p}(-i\partial_{x},k_{y})&=&\lambda_{y}k_{y}\sigma_{y}\tau_{z}+\Lambda(-i\partial_{x})\sigma_{x}s_{x}\tau_{z}+\Delta(-i\partial_{x})s_{y}\tau_{y},\quad
\end{eqnarray}
where  $\Lambda(-i\partial_{x})=\Lambda+\frac{1}{2}\Lambda_{x} \partial_{x}^{2}$
and $\Delta(-i\partial_{x})=\Delta+\frac{1}{2}\Delta_{x} \partial_{x}^{2}$. Here
we have neglected the insignificant $k_{y}^{2}$ terms for simplicity.
Treating $H_{p}$ as a perturbation and first solving the eigenvalue problem $H_{0}(-i\partial_{x})\psi_{\alpha}(x)=E_{\alpha}\psi_{\alpha}(x)$ under
the boundary condition $\psi_{\alpha}(0)=\psi_{\alpha}(+\infty)=0$, one can find that there are
four zero-energy solutions, which read
\begin{eqnarray}
\psi_{\alpha}(x)=\mathcal{N}_{x}\sin(\eta_{1}x) e^{-\eta_{2}x}e^{ik_{y}y}\chi_{\alpha},
\end{eqnarray}
where $\mathcal{N}_{x}=2\sqrt{|\eta_{2}(\eta_{1}^{2}+\eta_{2}^{2})/\eta_{1}^{2}|}$ denotes
the normalization factor, $\eta_{1}=\sqrt{2m/t_{x}-\lambda_{x}^{2}/t_{x}^{2}}$ and
$\eta_{2}=\lambda_{x}/t_{x}$; The four spinors $\chi_{\alpha}$ are determined by
$\sigma_{y}s_{z}\tau_{z}\chi_{\alpha}=-\chi_{\alpha}$. For their concrete forms, here
we follow ref.\cite{Yan2018hosc}. Accordingly, the matrix elements of $H_{p}$ under the basis composed
by the four zero-energy solutions are
\begin{eqnarray}
H_{{\rm I},\alpha\beta}(k_{y})=\int_{0}^{+\infty}\psi_{\alpha}^{\dag}(x)(-i\partial_{x},k_{y})\psi_{\beta}(x)dx.
\end{eqnarray}
The corresponding  low-energy Hamiltonian for edge (I) is
\begin{eqnarray}
H_{\rm I}(k_{y})=-\lambda_{y}k_{y}s_{z}+M_{{\rm I, \Lambda}}s_{y}+M_{{\rm I,S}}s_{y}\tau_{y},
\end{eqnarray}
where the two Dirac masses $M_{{\rm I, \Lambda}}$ and $M_{{\rm I,S}}$ are of different origins,
and they are given by
\begin{eqnarray}
M_{{\rm I,\Lambda}}&=&-\int_{0}^{+\infty}dx \psi_{\alpha}^{\dag} \Lambda(-i\partial_{x})\psi_{\alpha}(x)
=-\Lambda+\frac{m\Lambda_{x}}{t_{x}},\nonumber\\
M_{{\rm I,S}}&=&\int_{0}^{+\infty}dx \psi_{\alpha}^{\dag} \Delta(-i\partial_{x})\psi_{\alpha}(x)
=\Delta-\frac{m\Delta_{x}}{t_{x}}.
\end{eqnarray}
Similarly, the  low-energy Hamiltonians for the other three edges are
\begin{eqnarray}
H_{\rm II}(k_{x})&=&\lambda_{x}k_{x}s_{z}+M_{{\rm II,\Lambda}}s_{y}+M_{{\rm II,S}}s_{y}\tau_{y},\nonumber\\
H_{\rm III}(k_{y})&=&\lambda_{y}k_{y}s_{z}+M_{{\rm III,\Lambda}}s_{y}+M_{{\rm III,S}}s_{y}\tau_{y},\nonumber\\
H_{\rm IV}(k_{x})&=&-\lambda_{x}k_{x}s_{z}+M_{{\rm IV,\Lambda}}s_{y}+M_{{\rm IV,S}}s_{y}\tau_{y},
\end{eqnarray}
with $M_{{\rm II,\Lambda}}=M_{{\rm IV,\Lambda}}=-\Lambda-m\Lambda_{y}/t_{y}$,
$M_{{\rm II,S}}=M_{{\rm IV,S}}=\Delta-m\Delta_{y}/t_{y}$, and
$M_{{\rm III,\Lambda}}=M_{{\rm I,\Lambda}}$, $M_{{\rm III,S}}=M_{{\rm I,S}}$. By using
the boundary coordinate, the low-energy Hamiltonian can be written compactly as
\begin{eqnarray}
H_{\rm Edge}=-i\lambda(l)\partial_{l}s_{z}+ M_{\rm \Lambda}(l)s_{y}+M_{{\rm S}}(l)s_{y}\tau_{y},
\end{eqnarray}
where $\lambda(l)$,  $M_{\rm \Lambda}(l)$ and $M_{{\rm S}}(l)$ are step functions with
their values following the sequences: $\lambda(l)=\lambda_{x}$, $\lambda_{y}$, $\lambda_{x}$, $\lambda_{y}$,
$M_{\rm \Lambda}(l)=-\Lambda+m\Lambda_{x}/t_{x}$, $-\Lambda-m\Lambda_{y}/t_{y}$, $-\Lambda+m\Lambda_{x}/t_{x}$, $-\Lambda-m\Lambda_{y}/t_{y}$,
and $M_{\rm S}(l)=\Delta-m\Delta_{x}/t_{x}$, $\Delta-m\Delta_{y}/t_{y}$, $\Delta-m\Delta_{x}/t_{x}$, $\Delta-m\Delta_{y}/t_{y}$
for (I), (II), (III) and (IV), respectively.

Without loss of generality, let us focus on the case with $\Lambda_{x}=\Lambda_{y}$ so that $\Lambda=0$.
In the absence of pairing, i.e., $M_{{\rm S}}(l)=0$, $H_{\rm Edge}$ reduces
to a $2\times2$ matrix. At each corner,  $\lambda(l)$ does not change sign,
but $M_{\rm \Lambda}(l)$ does, realizing a domain wall of Dirac mass
which harbors one charged zero mode according to the Jackiw-Rebbi theory\cite{jackiw1976b}.
When superconductivity enters, one can see that
$H_{\rm Edge}$ is the direct sum of two independent parts, i.e., $H_{\rm Edge}=H_{\tau_{y}=1}\oplus H_{\tau_{y}=-1}$ with
\begin{eqnarray}
H_{\tau_{y}=1}&=&-i\lambda(l)\partial_{l}s_{z}+ (M_{\rm \Lambda}(l)+M_{{\rm S}}(l))s_{y},\nonumber\\
H_{\tau_{y}=-1}&=&-i\lambda(l)\partial_{l}s_{z}+ (M_{\rm \Lambda}(l)-M_{{\rm S}}(l))s_{y}.\label{domain}
\end{eqnarray}
One can see that the Dirac mass induced by superconductivity takes different signs in
the two parts. In the weak-pairing limit, $|M_{{\rm S}}(l)|<<|M_{\rm \Lambda}(l)|$, each part realizes
one zero mode per corner. As the particle component and the hole component of these zero modes' wave functions
are equal (note $\tau_{y}\psi_{0}(l)=\pm\psi_{0}(l)$, where $\psi_{0}(l)$ denotes the wave function of zero mode),
they are MZMs, agreeing with our previous argument that weak superconductivity will
transform one charged zero mode to two MZMs. As now each corner harbors two MZMs,
these MCMs are not stable. Indeed, we find that any finite $\mu$
or on-site potential will make them couple (the chemical potential term contains $\tau_{z}$,
so it makes the $\tau_{y}=1$ part couple with the $\tau_{y}=-1$ part) and consequently destroy their self-conjugate nature.
This can also be understood from the perspective that because $\mu$ shifts the energy of charged corner modes
away from zero, the energy of corner modes will keep taking finite values if the superconductivity is very weak.
Therefore, for the square geometry presented in Fig.\ref{sketch}, robust MCMs are absent in the weak-pairing limit.
Noteworthily, as $M_{\rm \Lambda}(l)$ is in fact sensitive to the orientation of edge, here we have emphasized the
particular square geometry shown in Fig.\ref{sketch}. As will see shortly, if the sample's geometry
is appropriately designed, the critical value of pairing amplitude for realizing robust MCMs can be
very small, so even weak superconductivity is sufficient.

To see how robust MCMs emerge in a square sample, we take $s$-wave pairing for illustration
(other more exotic cases can similarly be analyzed).
Accordingly, $M_{\rm S}(l)=\Delta_{0}$ is uniform on the boundary.
Without loss of generality,  we further assume $m\Lambda_{y}/t_{y}>m\Lambda_{x}/t_{x}$.
According to Eq.(\ref{domain}),
one can find when $m\Lambda_{x}/t_{x}<\Delta_{0}<m\Lambda_{y}/t_{y}$, while the domain walls for
$H_{\tau_{y}=1}$ are preserved since $m\Lambda_{x}/t_{x}+\Delta_{0}$ and  $-m\Lambda_{y}/t_{y}+\Delta_{0}$
still take opposite signs, the ones for $H_{\tau_{y}=-1}$ are removed since now
$m\Lambda_{x}/t_{x}-\Delta_{0}$ and  $-m\Lambda_{y}/t_{y}-\Delta_{0}$ take
same sign. As a result,  there is only one MZM per corner in this regime, as
shown in Fig.\ref{twomcm}(a).  We have numerically checked that these MCMs are robust against local perturbations,
doping and random disorders as long as the doping level and disorder strength are small
than some critical values (note in Fig.\ref{twomcm}(a), $\mu=0.1$).

According to the criterion $m\Lambda_{x}/t_{x}<\Delta_{0}<m\Lambda_{y}/t_{y}$, one may make
the conclusion that if the underlying pairing is $s$-wave, anisotropy
is necessary for the realization of SOTSC. That is, if $\Lambda_{x}=\Lambda_{y}$,
$t_{x}\neq t_{y}$ must be satisfied. However, anisotropy is in fact unnecessary.
For the isotropic case with $t_{x}=t_{y}$ and $\Lambda_{x}=\Lambda_{y}$,
$M_{\rm \Lambda}(l)$ follows the angle dependence $M_{\rm \Lambda}(l)\simeq m\Lambda_{x}\cos2\theta/t_{x}$,
where $\theta$ represents the angle relative to edge (I). This indicates that on the edge
whose orientation is pointing to $\theta=\pi/4$, $M_{\rm \Lambda}(l)=0$. As a result,  one can find that for the
$\pi/4$-angle corner
formed by edge (I) and the $\theta=\pi/4$-orientation edge, it will harbor one MZM as long as
$0<|\Delta_{0}|<m\Lambda_{x}/t_{x}$. We demonstrate the validity of this analysis numerically, as shown in
Figs.\ref{twomcm}(b)(c).
According to the phase diagram in Fig.\ref{twomcm}(c), one can
see that for an isosceles-right-triangle geometry, MCMs can exist for a quite broad range of $\mu$ and for
infinitely weak pairing amplitude\cite{supplemental}.

As for a SOTI, $M_{\rm \Lambda}(l)$ inevitably vanishes along some direction, this implies
that  a judicious
design of the corners is always able to realize MCMs even though the superconductivity is weak. Clearly,
this conclusion also holds for other unconventional SCs.

\begin{figure}
\subfigure{\includegraphics[width=4.25cm, height=4cm]{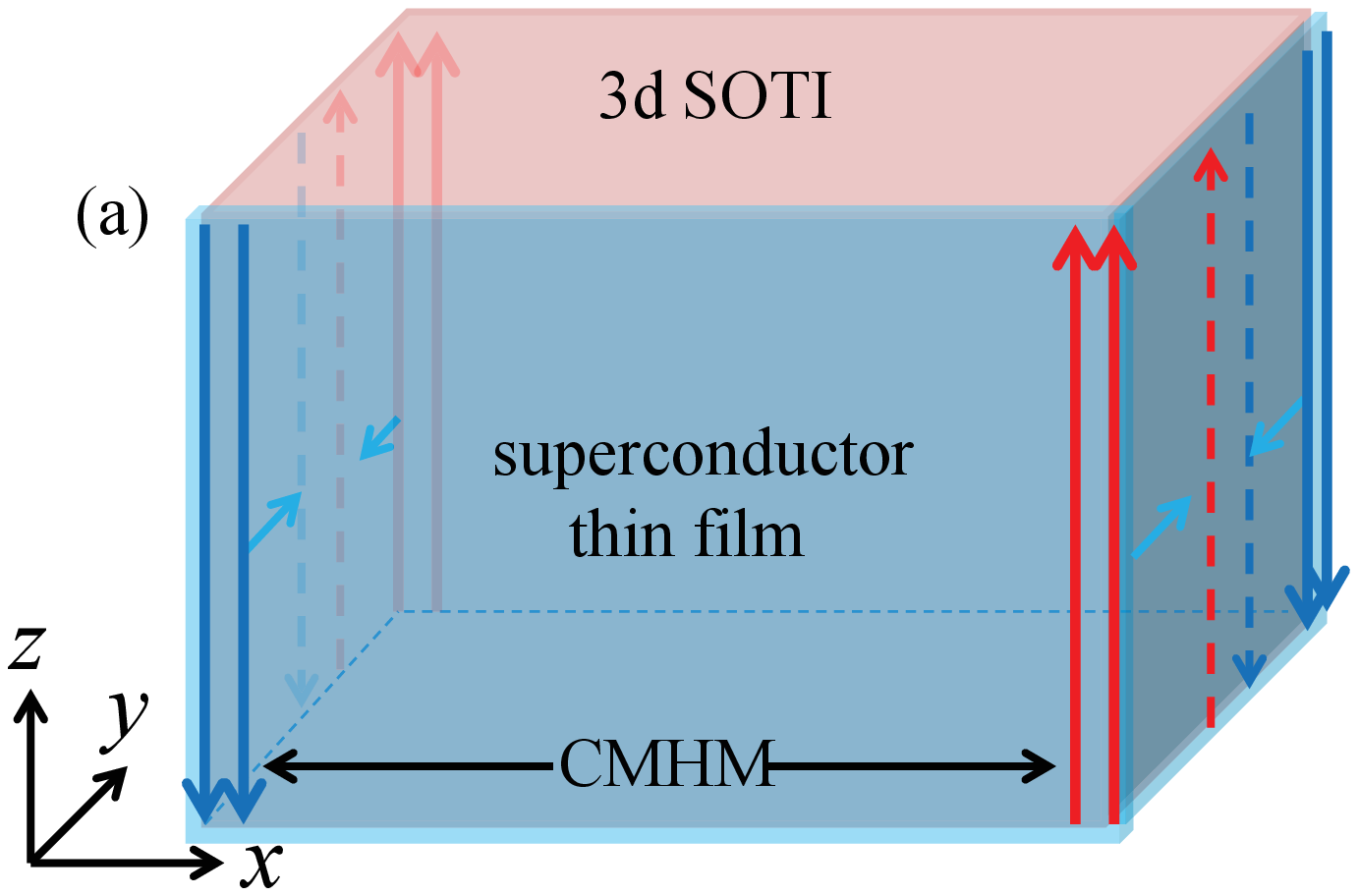}}
\subfigure{\includegraphics[width=4.25cm, height=4cm]{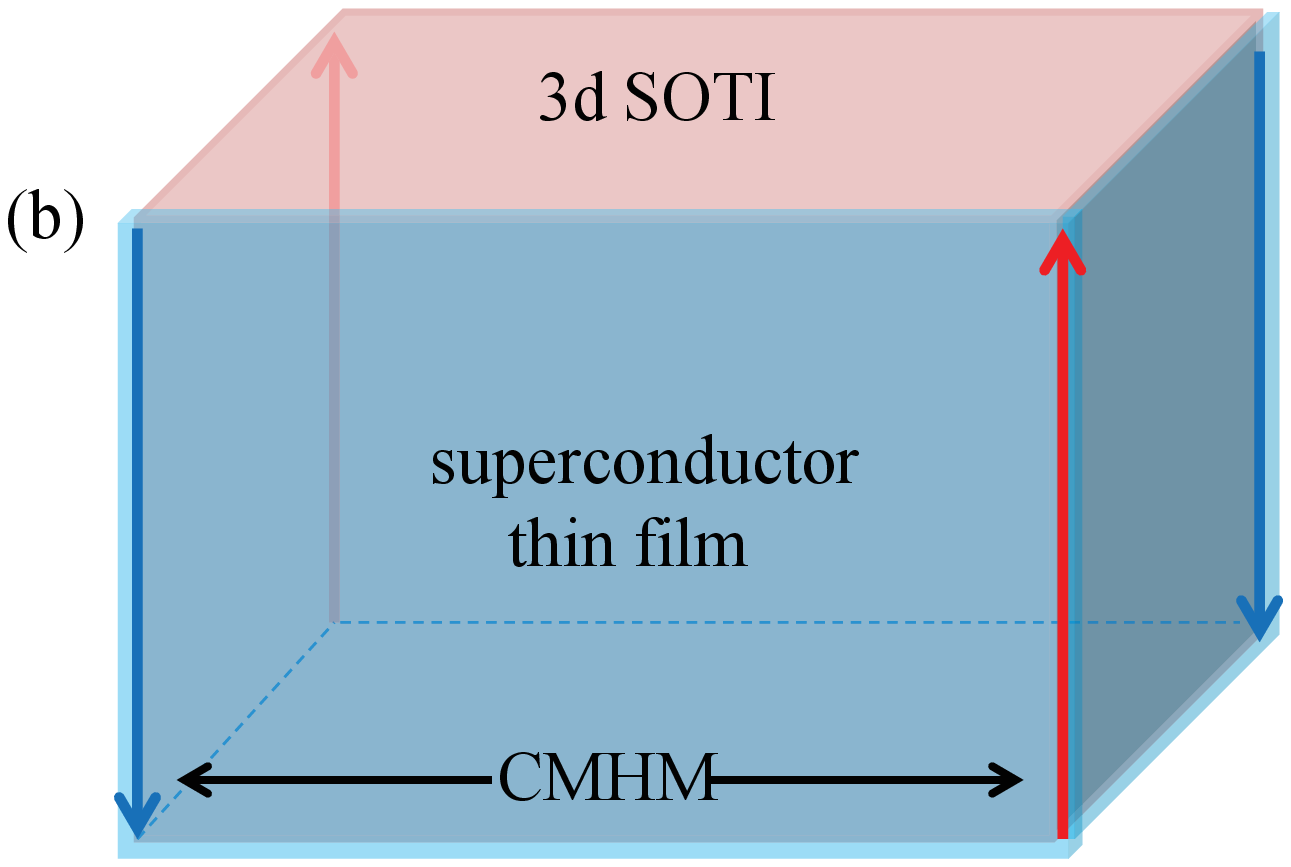}}
\subfigure{\includegraphics[width=4.25cm, height=4cm]{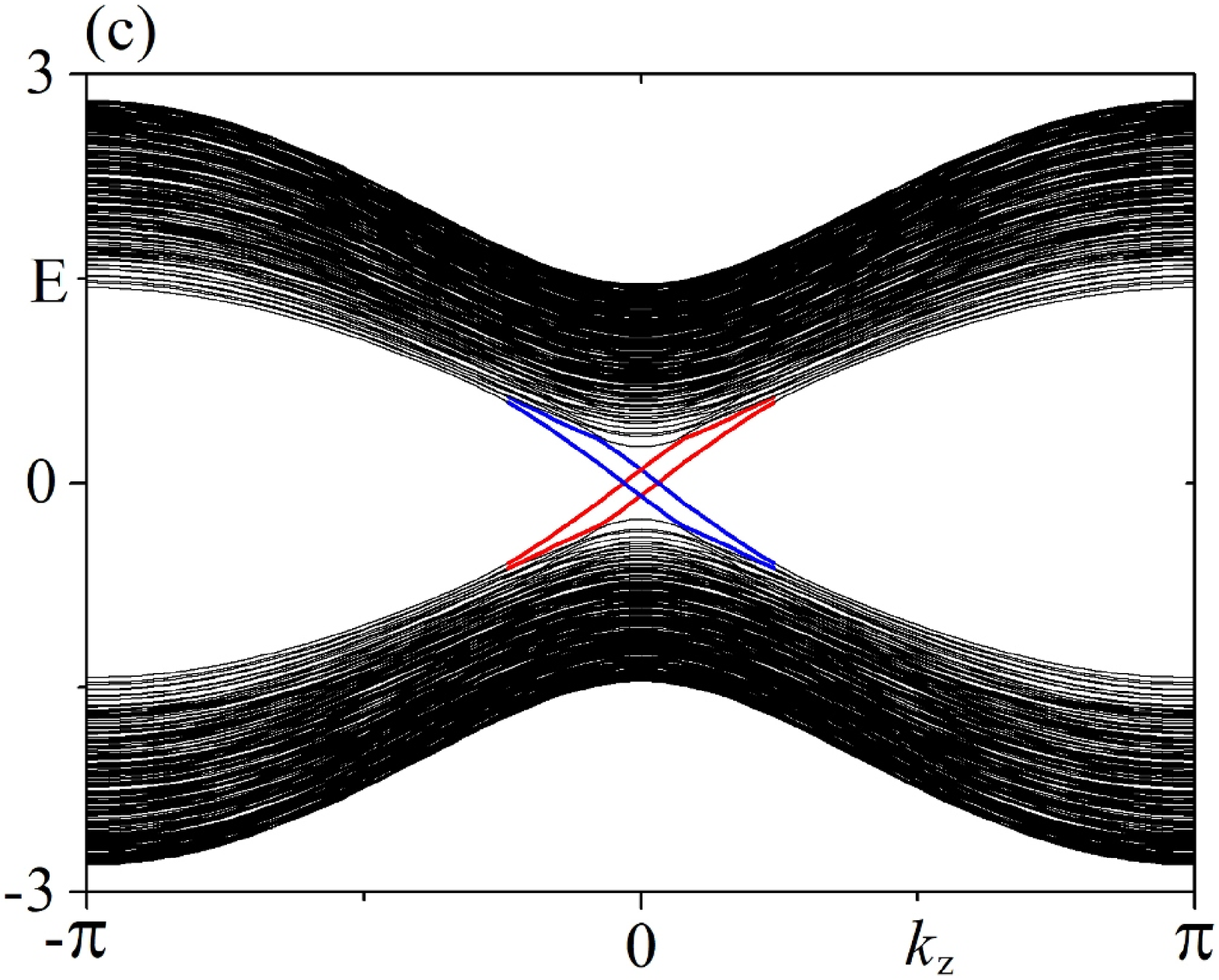}}
\subfigure{\includegraphics[width=4.25cm, height=4cm]{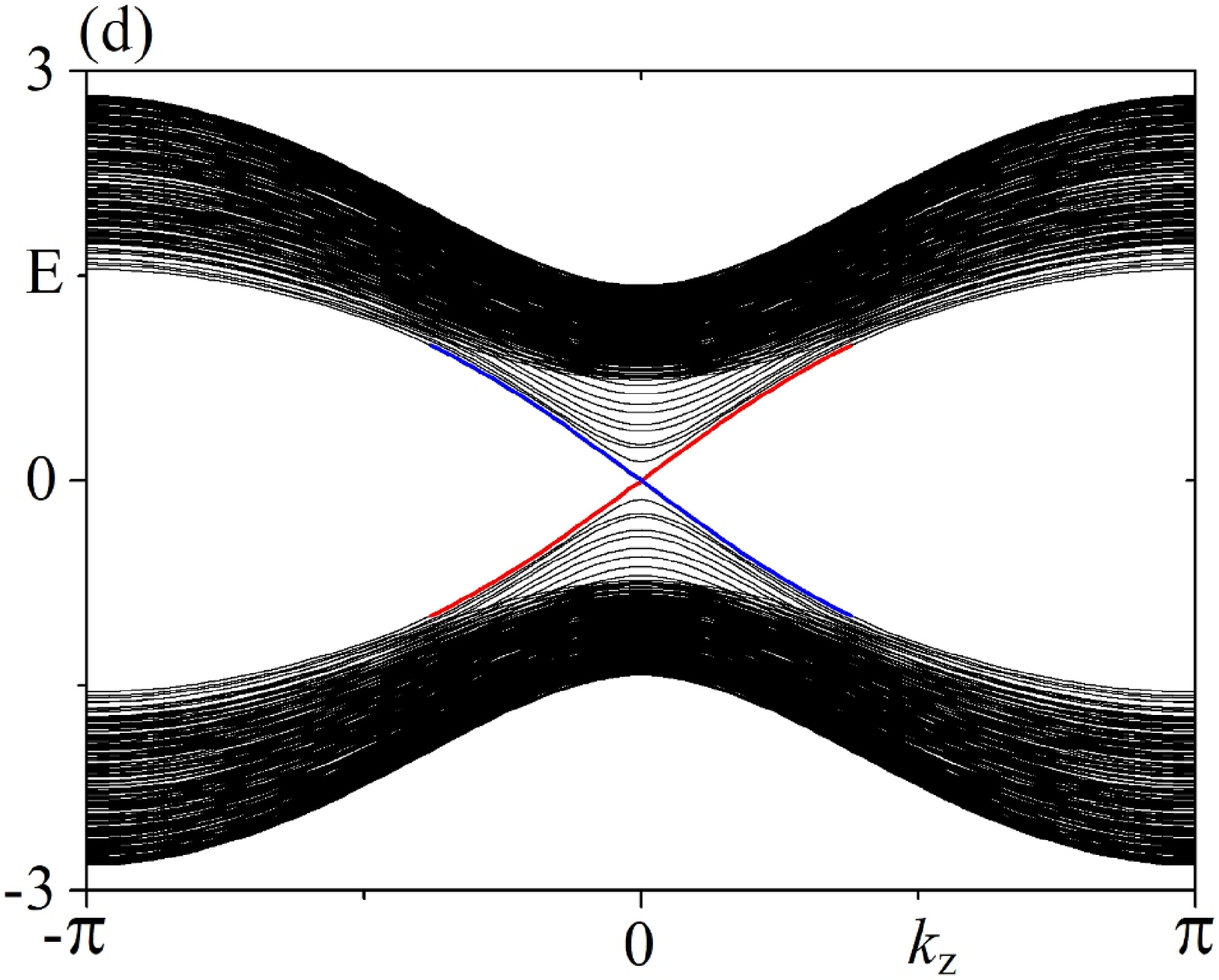}}
\caption{ CMHMs in a $3d$ SOTI/s-wave SC heterostructure. Common parameters: $t_{x}=t_{z}=1.0$, $t_{y}=0.5$, $m_{0}=2.0$,
$\lambda_{x}=\lambda_{y}=\lambda_{z}=\Lambda_{x}=\Lambda_{y}=1.0$, and $\mu=0.1$. (a)(b) Schematic illustration of the topological phase transition
between two SOTSC phases. In the weak-pairing limit, there are two chiral Majorana modes per hinge.
With the increase of pairing amplitude, four of the eight chiral Majorana modes will move towards each other in pairs (see the dashed
lines in (a)) and then annihilate, resulting in a new SOTSC phase with one chiral Majorana mode per hinge.
(c) and (d) show the energy spectra for a geometry with $L_{x}\times L_{y}=20\times 20$ and periodic boundary condition in
the $z$ direction. (c) $\Delta_{0}=0.1$, (d)  $\Delta_{0}=0.6$. The red lines traversing the gap represent up-moving CMHMs, and the blue lines
traversing the gap represent down-moving CMHMs. All red and blue lines are doubly-degenerate. The number and the distributions of up-moving and
down-moving CMHMs in (c) and (d) are consistent with the pictures presented in (a) and (b), respectively.
} \label{three}
\end{figure}

{\it CMHMs in a $3d$ SOTI/SC heterostructure.---} The scenario above can straightforwardly be
generalized to $3D$. For example, if we have a $3d$ SOTI at hand, we can grow a thin film of
$s$-wave SC on its surface (see Figs.\ref{three}(a)(b)). Accordingly, the system could be modeled by
$H=\sum_{\bk}\Psi_{\bk}^{\dag}H(\bk)\Psi_{\bk}$
with
\begin{eqnarray}
H(\bk)&=&\xi(\bk)\sigma_{z}\tau_{z}+\sum_{i=x,z}\lambda_{i}\sin k_{i}\sigma_{x}s_{i}+\lambda_{y}\sin k_{y}\sigma_{x}s_{y}\tau_{z}\nonumber\\
&&+\Lambda(\bk)\sigma_{y}+\mu\tau_{z}+\Delta_{0}s_{y}\tau_{y},
\end{eqnarray}
where $\xi(\bk)=m_{0}-t_{x}\cos k_{x}-t_{y}\cos k_{y}-t_{z}\cos k_{z}$. Without the terms in the second line,
the Hamiltonian describes a strong TI when $\prod_{\alpha,\beta,\gamma=\pm1}(m_{0}+\alpha t_{x}+\beta t_{y}
+\gamma t_{z})<0$. Accordingly, when open boundary condition is taken, gapless Dirac surface states
will appear on the boundary.  The presence of the term $\Lambda(\bk)$ gaps out the Dirac surface states on the four lateral surfaces (in $3D$, we
take open boundary condition in both the $x$ and $y$ directions, and periodic boundary condition in the $z$ direction) and
leaves one chiral electronic mode per hinge\cite{Schindler2018HOTIa}.

As mentioned before, when superconductivity enters, one chiral electronic mode becomes two chiral Majorana
modes in the weak-pairing limit\cite{qi2010chiral}. Unlike the MCMs in two dimensions, while here the wave functions of the two chiral Majorana modes also overlap in space,
they are stable against perturbations since they are chiral in nature. Therefore, in the weak-pairing regime,
there are two robust chiral Majorana modes per hinge (see Figs.\ref{three}(a)(c)). Interestingly, we find that with the increase of pairing amplitude,
a topological phase transition will take place on the boundary and accordingly a new SOTSC which host one robust chiral
Majorana mode per hinge will be realized (see  Figs.\ref{three}(b)(d)). It is worth noting that when
doing the calculation of the energy spectra presented in Figs.\ref{three}(c)(d), the superconductivity has
been taken to be uniform throughout the whole sample. It is apparent that this assumption is unrealistic
for the heterostructure since deep in the bulk the superconductivity induced by proximity effect should vanish, however,
the low-energy physics within the gap can be well captured
since the in-gap states are located on the surfaces which are well in contact with the SC.
In other words, here the derivation from real situation only has strong impact on the bulk states. In fact, if
we focus on the in-gap states, we can also adopt the edge theory as in $2D$.  For the geometry shown in Figs.\ref{three}(a)(b),
one can easily find that the criterion for realizing the SOTSC phase with one robust chiral
Majorana mode per hinge is also
$m\Lambda_{x}/t_{x}<\Delta_{0}<m\Lambda_{y}/t_{y}$
($m=t_{x}+t_{y}+t_{z}-m_{0}>0$, $\mu=0$, and $\Lambda_{x}=\Lambda_{y}$ are also presumed).
One can see that the results presented in Figs.\ref{three}(c)(d) are consistent with this criterion.
Similar to the $2d$ situation,
the critical pairing amplitude can also be tuned to take a very small value if the sample's geometry
is appropriately designed.

{\it Conclusions.---} We have shown that SOTI/SC heterostructures provide
promising new platforms of MCMs and CMHMs.  As our proposed scheme requires
neither special pairings nor magnetic fields, we believe it should be
simple to implement experimentally.  Consider the fast growth of material candidates
for SOTIs\cite{schindler2018HOTI,yue2019symmetry,Wang2018XTe,Xu2019HOTI,Sheng2019SOTI,Lee2019HOTI,Chen2019HOTI},
we can foresee that such novel heterostructures  will be synthesised and
investigated in the near future. Experimentally,  MCMs and CMHMs can be probed by
STM techniques\cite{Jack2019observation} and transport experiments\cite{Gray2019helical}.

{\it Acknowlegements.---} We would like to acknowledge the support
by a startup grant at Sun Yat-sen University.

\bibliography{dirac}

\vspace{8mm}

\newpage

{\bf Supplemental Material}

\vspace{4mm}

In this supplemental material, we provide the details about the determination of the phase diagram.
Let us first rewrite down the Hamiltonian, which is
\begin{eqnarray}
H(\bk)&=&\epsilon(\bk)\sigma_{z}\tau_{z}+\lambda_{x}\sin k_{x}\sigma_{x}s_{z}+\lambda_{y}\sin k_{y}\sigma_{y}\tau_{z}\nonumber\\
&&+\Lambda(\bk)\sigma_{x}s_{x}\tau_{z}+\mu\tau_{z}+\Delta(\bk)s_{y}\tau_{y},\label{model}
\end{eqnarray}
where $\sigma_{i}$, $s_{i}$ and $\tau_{i}$ are Pauli matrices
in orbit $(a,b)$, spin ($\uparrow,\downarrow$) and particle-hole spaces, respectively;
$\epsilon(\bk)=m_{0}-t_{x}\cos k_{x}-t_{y}\cos k_{y}$ is the kinetic energy;
$\Lambda(\bk)=\Lambda_{x}\cos k_{x}-\Lambda_{y}\cos k_{y}$ is a time-reversal symmetry breaking term
crucial for the realization of SOTI;
$\mu$ is the chemical potential, and $\Delta(\bk)=\Delta_{0}+\Delta_{x}\cos k_{x}+\Delta_{y}\cos k_{y}$ represents
the pairing. Here we focus on conventional $s$-wave superconductor, so we let $\Delta_{x}=\Delta_{y}=0$.

The Hamiltonian has an intrinsic particle-hole symmetry, i.e., $PH(\bk)P^{-1}=-H(-\bk)$
with $P=\tau_{x}K$ ($K$ denotes the charge conjugate). For the special case with $\mu=0$,
the Hamiltonian has an additional chiral symmetry,
i.e., $\{C, H(\bk)\}=-H(\bk)$ with $C=\sigma_{x}s_{y}\tau_{z}$.
The energy spectra of this Hamiltonian are given by
\begin{eqnarray}
E(\bk)=\pm\sqrt{F(k)\pm 2\sqrt{G(k)}},
\end{eqnarray}
where $F(k)=\epsilon^{2}(\bk)+\lambda_{x}^{2}\sin^{2} k_{x}+\lambda_{y}^{2}\sin^{2} k_{y}
+\Lambda^{2}(\bk)+\Delta_{0}^{2}+\mu^{2}$, and $G(k)=\mu^{2}(\epsilon^{2}(\bk)+\lambda_{x}^{2}\sin^{2} k_{x}+\lambda_{y}^{2}\sin^{2} k_{y}
+\Lambda^{2}(\bk))+\Lambda^{2}(\bk)\Delta_{0}^{2}$.
If without the terms in the second line the Hamiltonian in Eq.(\ref{model}) describes an insulator,
then the above energy spectra are always gapped as long as $\Delta_{0}\neq 0$.

\begin{figure}
\subfigure{\includegraphics[width=4.25cm, height=4cm]{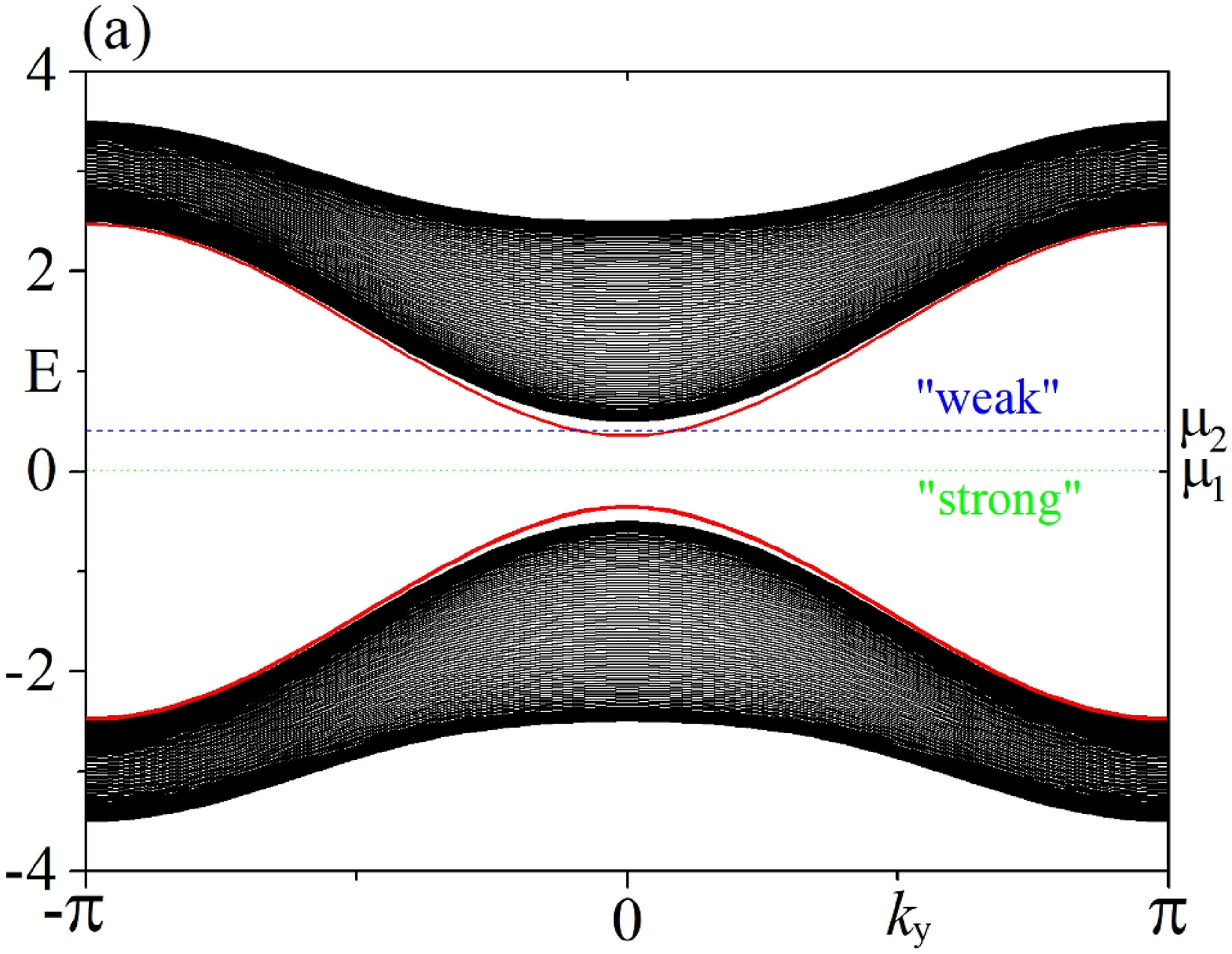}}
\subfigure{\includegraphics[width=4.25cm, height=4cm]{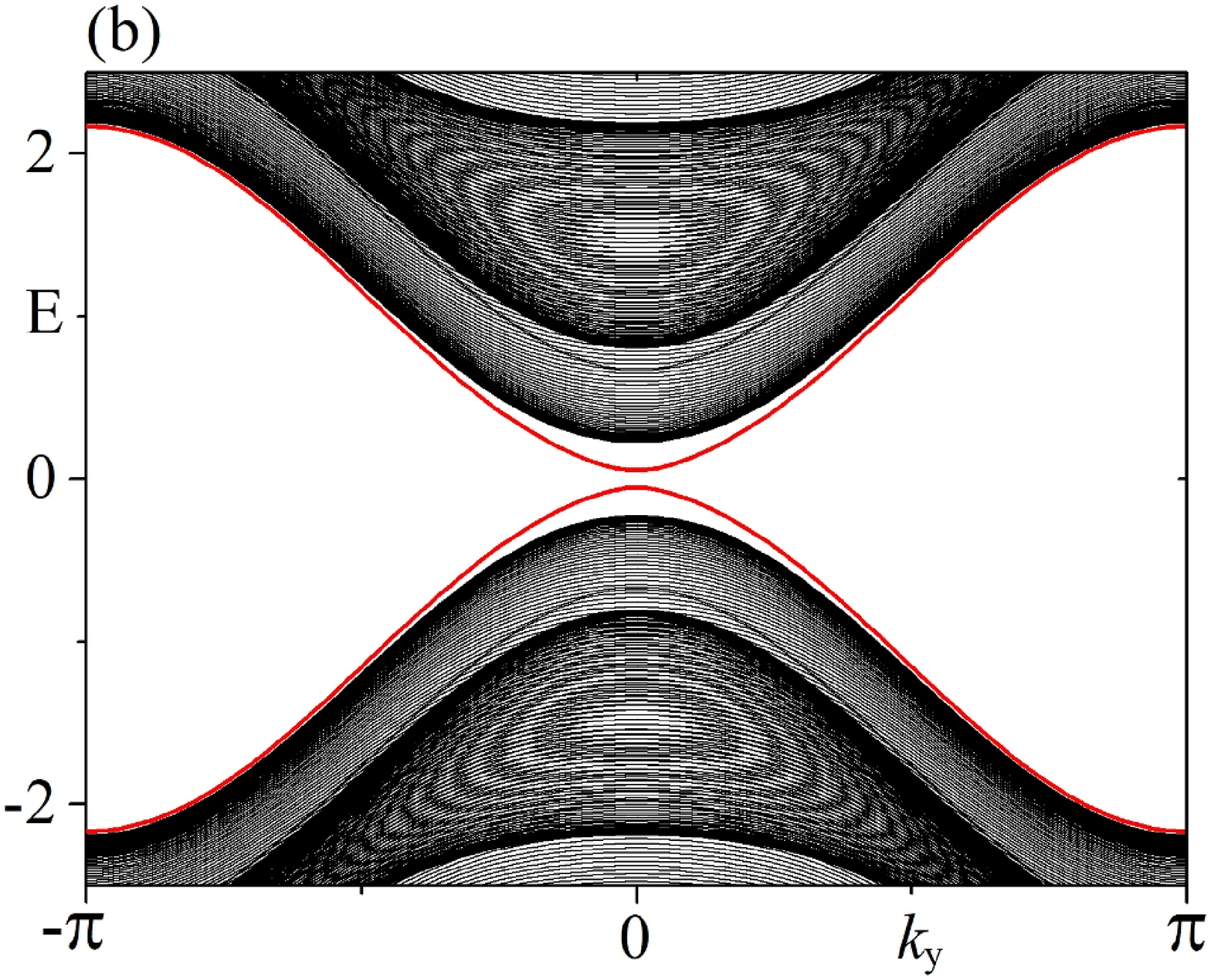}}
\subfigure{\includegraphics[width=4.25cm, height=4cm]{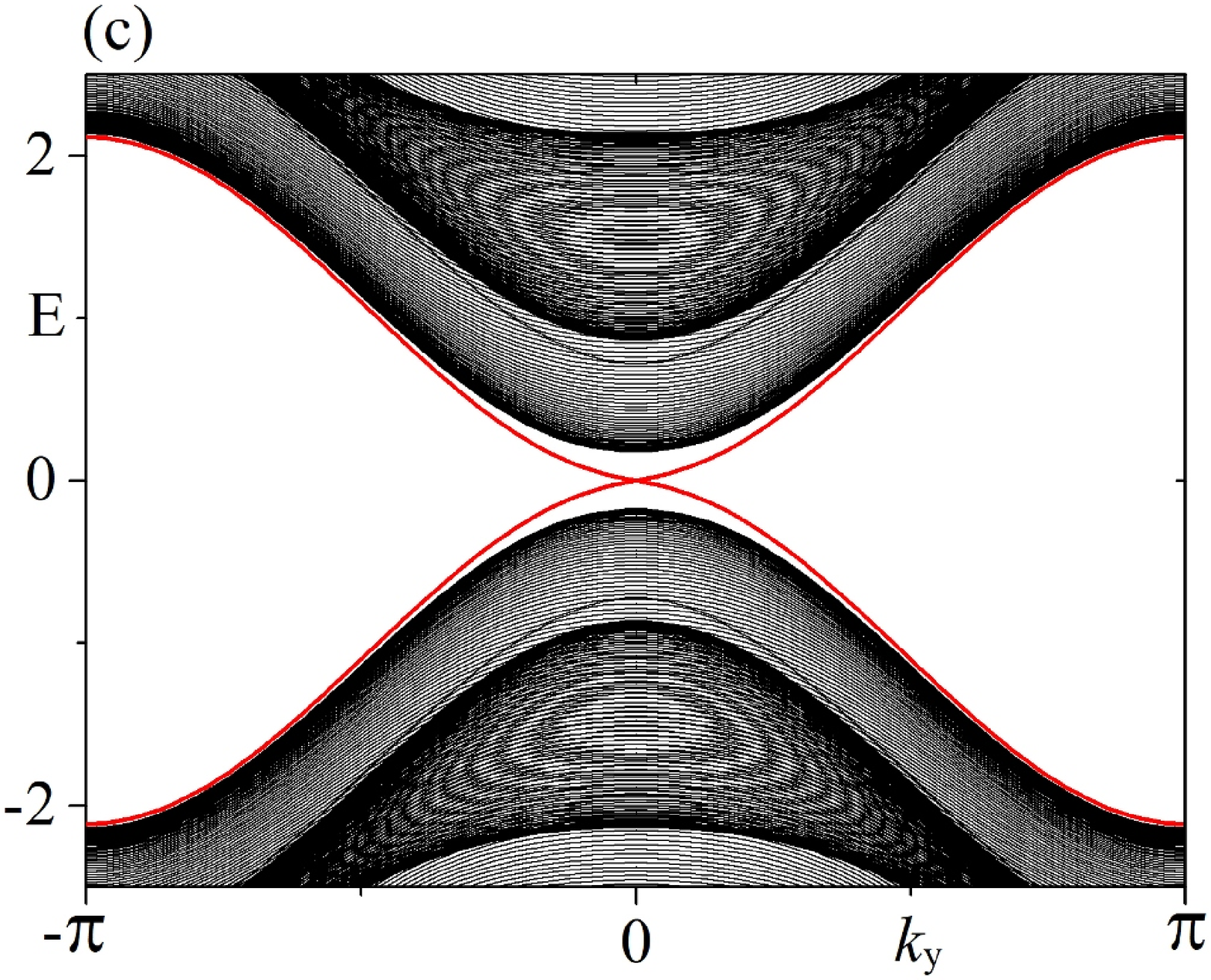}}
\subfigure{\includegraphics[width=4.25cm, height=4cm]{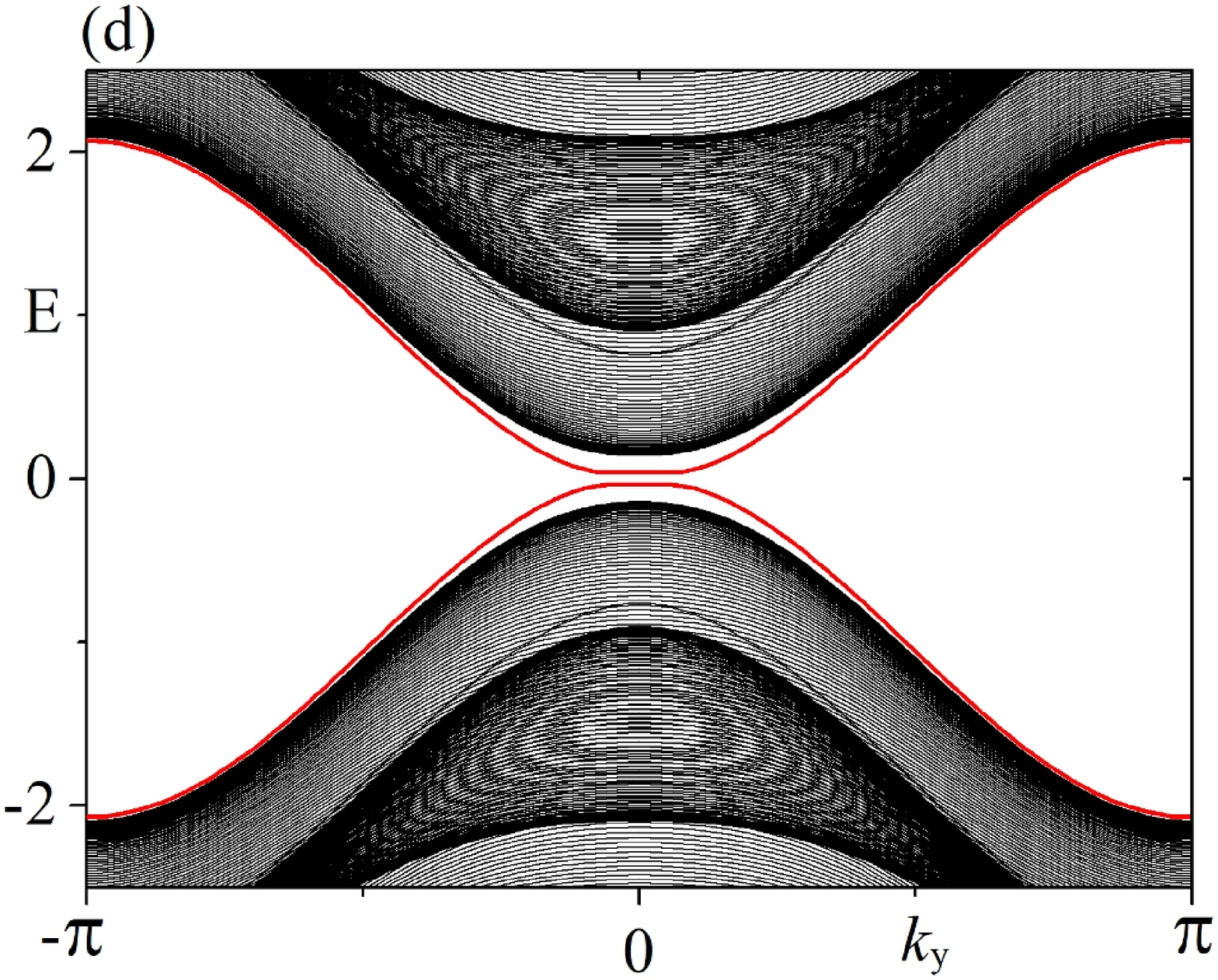}}
\caption{Energy spectra for a sample which takes open boundary condition in the $x$ direction, and
periodic boundary condition in the $y$ direction. The lattice size along the $x$ direction is $L_{x}=100$.
The red lines denote the energy spectra of edge modes.
Common parameters are $t_{x}=t_{y}=\lambda_{x}=\lambda_{y}=\Lambda_{x}=\Lambda_{y}=1$, and $m_{0}=1.5$. (a)
The superconductivity is absent, i.e., $\Delta_{0}=0$. When $\mu<E_{g}/2$, where $E_{g}$ denotes
the energy gap of edge-mode energy spectra, the chemical potential does not cross any spectra.
In (b)(c)(d), $\Delta_{0}=0.1$. (b) $\mu=0.3<\mu_{c}=0.356$; (c) $\mu=\mu_{c}$; (d) $\mu=0.4>\mu_{c}$.
One can see that $E_{g}$ undergoes an ``open-to-closed-to-open'' transition
with the increase of $\mu$ from zero.
  }\label{gap}
\end{figure}

Now we consider that without the terms in the second line in Eq.(\ref{model}), the Hamiltonian describes
a first-order topological insulator with gapless helical edge modes on the boundary. As mentioned in the main text, adding
the $\Lambda(\bk)$ term drives the system to a second-order topological insulator with localized corner modes. When
superconductivity enters, as bulk energy spectra keep gapped no matter what value the pairing amplitude and the chemical
potential take, this implies that the first-order topological property is always trivial.

The change of topological property (or say topological phase transition) is  associated with the close of energy gap. For a first-order topological phase, topological phase transition is associated with the close of bulk energy gap. Accordingly, for an $n$th-order topological phase in $d$ dimensions,
the topological phase transition is associated with the close of energy gap of the $(d-n+1)$-dimensional  boundary modes.
Guided by this principle, in the following we investigate the phase diagram.

\begin{figure}[t]
\subfigure{\includegraphics[width=4.25cm, height=4cm]{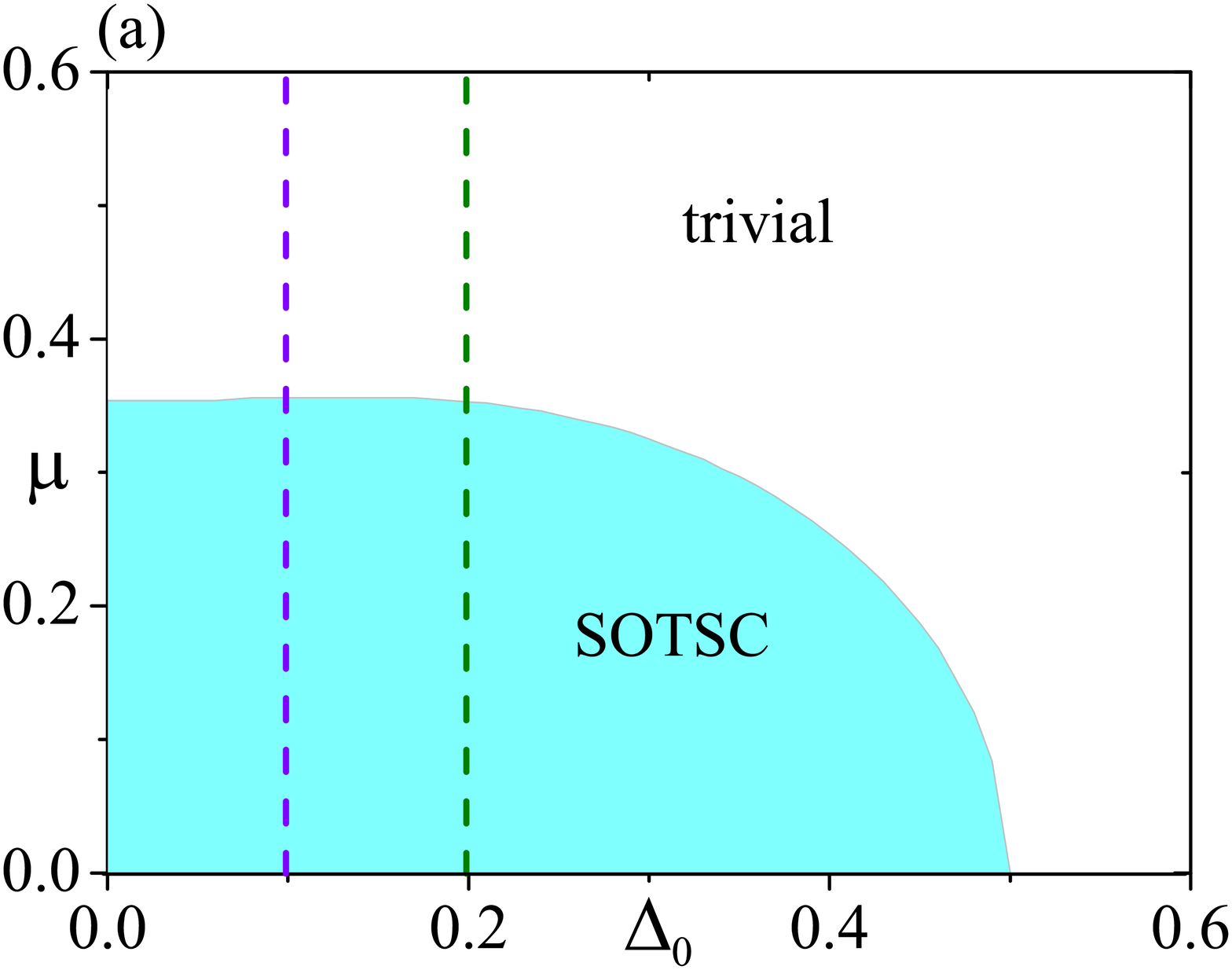}}
\subfigure{\includegraphics[width=4.25cm, height=4cm]{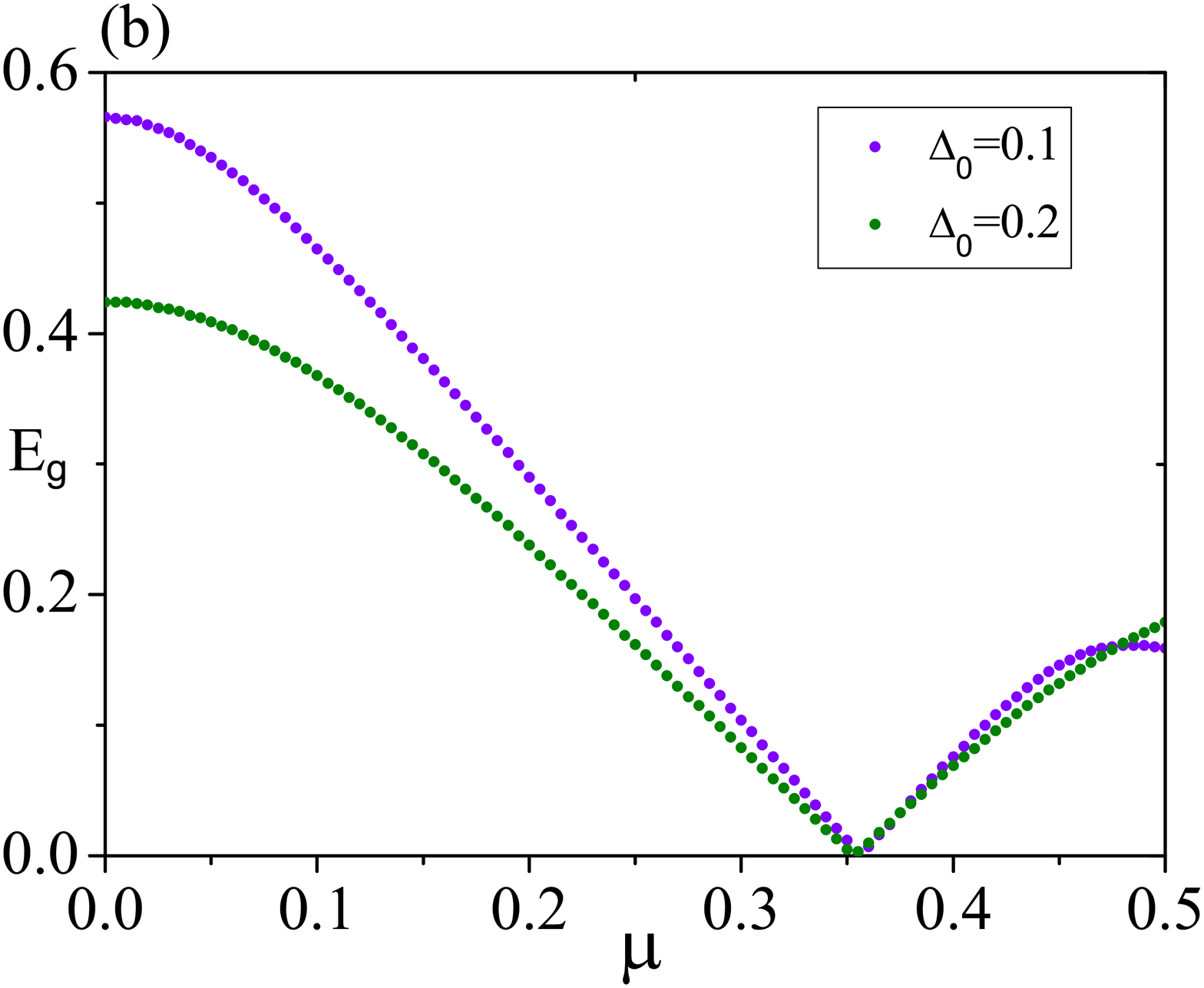}}
\subfigure{\includegraphics[width=4.25cm, height=4cm]{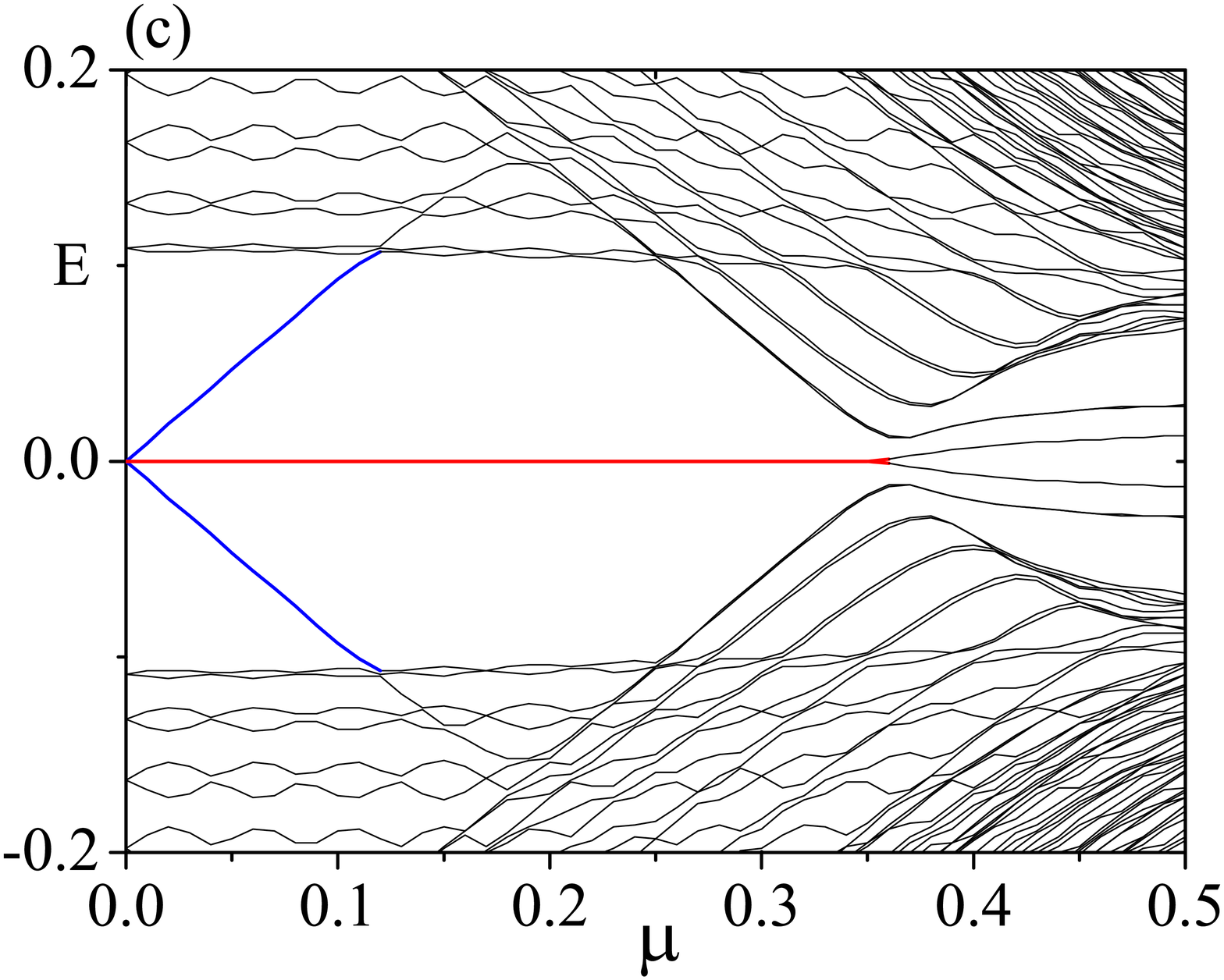}}
\subfigure{\includegraphics[width=4.25cm, height=4cm]{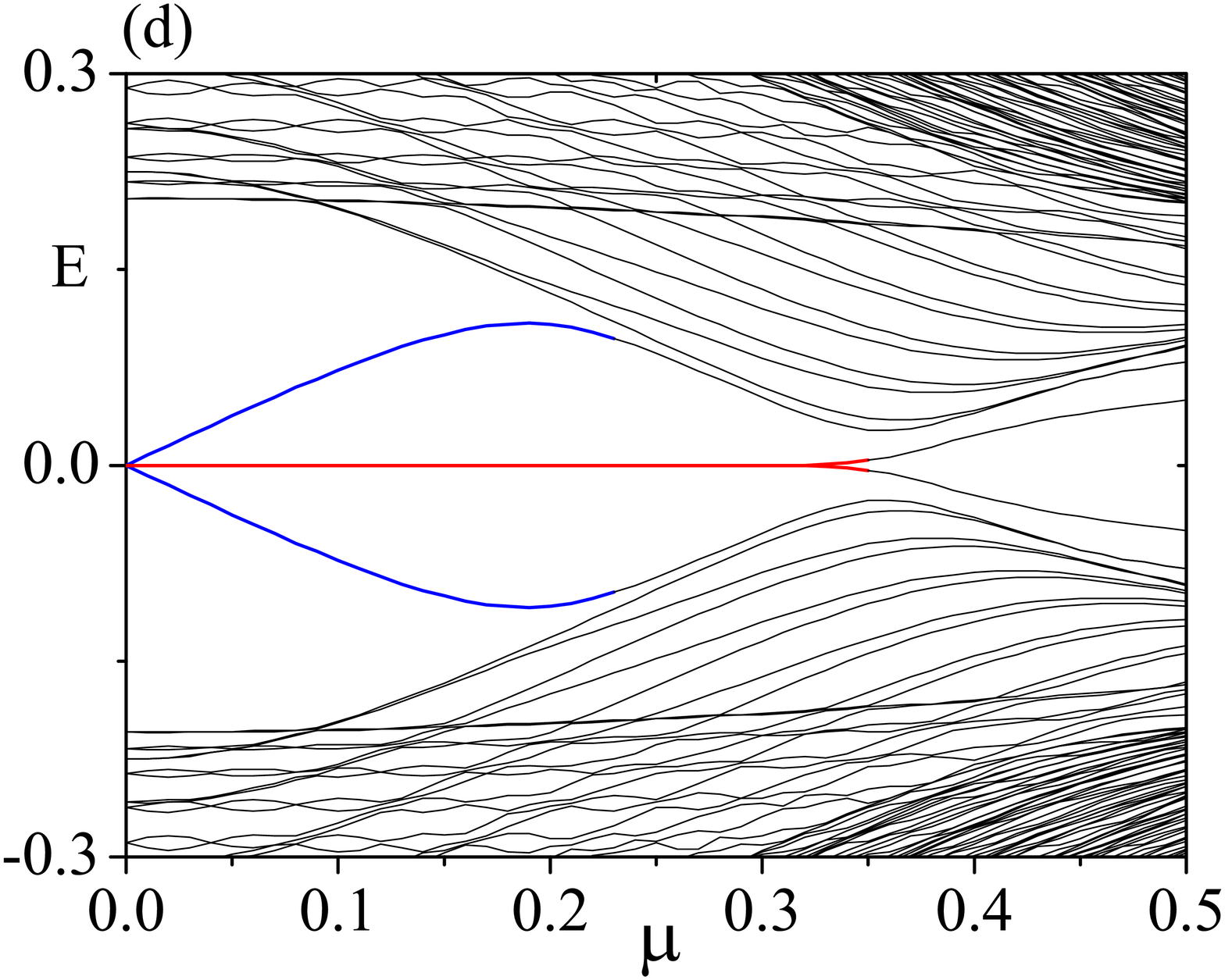}}
\caption{(a) Phase diagram for an  isosceles-right-triangle geometry. Common parameter are
$t_{x}=t_{y}=\lambda_{x}=\lambda_{y}=\Lambda_{x}=\Lambda_{y}=1$, and $m_{0}=1.5$. To show that
the phase diagram can be determined by simply investigating the  gap of
edge-mode energy spectra, we focus on the two dashed lines shown in (a) for
illustration.  (b) $E_{g}$-vs-$\mu$. For
a fixed pairing amplitude, with the increase of $\mu$, the  gap of edge-mode energy spectra for
the edge with orientation in parallel to the $y$ direction will undergo an ``open-to-closed-to-open'' transition.
Accordingly, the topological property of this edge undergoes a transition from the strong pairing regime to weak pairing
regime. (c)(d) Energy spectra
for an isosceles-right-triangle sample whose  length of the two right-angle sides are equal to $40$.
Here only the part near zero energy has been shown.  (c) $\Delta_{0}=0.1$; (d) $\Delta_{0}=0.2$.
In (c)(d), the red lines correspond to the energy spectra of the two Majorana corner modes. The blue lines correspond to
the energy spectra of two bound states located at the right-angle corner, one can see that once $\mu$ goes away from
zero, their energies are split. One can infer from (b)(c)(d) that
using the gap close of edge-mode energy spectra can faithfully determine the phase boundary.}\label{phase}
\end{figure}

For concreteness, we consider the isosceles-right-triangle geometry (see Fig.2(b) in the main text) and focus on the case with isotropic
parameters, i.e., $t_{x}=t_{y}=\lambda_{x}=\lambda_{y}=\Lambda_{x}=\Lambda_{y}=1$. Let us first focus on
the edge whose orientation is in parallel to the $y$ direction. To obtain the corresponding
energy spectra of edge modes, it is more convenient to consider that the system takes
open boundary condition in the $x$ direction and periodic boundary condition in the $y$ direction.
We first consider the case without superconductivity. As shown in Fig.\ref{gap}(a), the in-gap edge modes
are gapped, which is consistent with the trivialness of first-order topological property.
For convenience, we label the gap of edge-mode energy spectra as $E_{g}$. When $|\mu|<E_{g}/2$, there is no
boundary Fermi surface, therefore when superconductivity enters, the topological property of this edge corresponds to the strong pairing
regime\cite{read2000}. For a fixed pairing amplitude, with the increase of $\mu$, the boundary topological property will undergo
a transition from strong pairing regime to weak pairing regime\cite{read2000}. Accordingly, $E_{g}$ will undergo an ``open-to-closed-to-open'' transition (see Figs.\ref{gap}(b)(c)(d)). At the critical point,  it gets closed (see Fig.\ref{gap}(c)).
In comparison, as the $\Lambda(\bk)$ term vanishes along the
$k_{x}=k_{y}$ and $k_{x}=-k_{y}$ directions, the energy spectra of edge modes on the edge with
orientation pointing to $\theta=\pi/4$ ($\theta$ is defined in relative to the positive $y$ direction)
will keep gapless before the superconductivity enters, implying that the topological property on the $\theta=\pi/4$-orientation edge
always corresponds to the weak pairing regime.  As a result, when the topological property of
the $\theta=0$-orientation edge corresponds to the strong pairing regime,
the $\pi/4$-angle corner, which is the intersection of $\theta=0$-orientation edge and $\theta=\pi/4$-orientation edge, is a domain wall which harbors one
robust Majorana zero mode.
According to this principle, the phase diagram can be mapped out, as shown in Fig.\ref{phase}(a).
We have confirmed that the phase boundary determined by using this principle is consistent
with the approach  based on the direct diagonalization of the real-space Hamiltonian (see Fig.\ref{phase}(b)(c)(d)). Owing to the limitation in
computing power, we find that using the gap close of edge-mode energy spectra to
determine the phase boundary is much more efficient and precise than directly diagonalizing the
real-space Hamiltonian.

\end{document}